\documentclass[10pt]{article}
\pdfoutput=1
\usepackage{chicago,epic,graphpap,graphics,float,tabularx,paralist,longtable,fancyhdr,fancyvrb}
\usepackage{amsmath}
\usepackage{amstext}
\usepackage{amsbsy}
\usepackage{amsthm}
\usepackage{amsfonts}
\usepackage{amssymb}
\usepackage{amscd}
\usepackage{eucal}
\usepackage{bbm}
\numberwithin{equation}{section}
\newtheorem{thm}{Theorem}[section]

\theoremstyle{definition}
\newtheorem{defn}[thm]{Definition}

\newtheorem{rem}{Remark}

\def\g{\gamma}
\def\s{\sigma}
\def\a{\alpha}
\def\p{\pi}
\def\m{\mu}

\def\mtil{{\widetilde \mu}}

\def\bla{{\boldsymbol \lambda}}
\def\bsi{{\boldsymbol \sigma}}
\def\ph{\varphi}
\def\ito{It{\^o}}
\def\as{\quad\text{{\rm a.s.}}}

\def\eqdef{\triangleq}
\def\lt#1{\Lambda_{#1}}
\def\half{\frac{1}{2}}
\def\sumi{\sum_{i=1}^n}
\def\sumj{\sum_{j=1}^n}

\def\brac#1{\langle #1 \rangle}

\def\eps{\varepsilon}
\def\z{\zeta}

\def\nnu{{\boldsymbol \nu}}
\def\ppi{{\boldsymbol \pi}}
\def\aalpha{{\boldsymbol \alpha}}
\def\SSigma{{\boldsymbol\Sigma}}

\def\1{{\mathbbm 1}}
\def\intT{\int_0^T}

\def\sumij{\sum_{i,j=1}^n}
\def\limT#1{\lim_{T\to\infty}\frac{#1}{T}}

\title{\bf Numeraire markets}
\author{Robert Fernholz\thanks{Intech, One Palmer Square, Princeton, NJ 08542.  bob@bobfernholz.com. The author thanks Ioannis Karatzas and Johannes Ruf for their invaluable comments and suggestions regarding this research.}}
\date{\today}

\begin{document}
\maketitle

\begin{abstract} In a stock market, the {\em numeraire portfolio,} if it exists, is the portfolio with the highest expected logarithmic growth rate at all times. A {\em numeraire market} is a stock market for which the market portfolio is the numeraire portfolio. We study {\em open markets,} markets comprising the higher capitalization stocks within a broad equity universe. The stocks we consider are represented by continuous semimartingales, and we construct an example of a numeraire market that is  asymptotically stable.
\end{abstract}

\section{Introduction}%%%%%%%%%%%%%%%%%%%%%%%%%

The stock markets considered in mathematical finance are conventionally {\em closed markets,} which contain the same stocks at all times. Real stock markets are {\em open markets} in which stocks enter following IPOs or spinoffs and exit following privatizations or bankruptcies. Open markets are similar to high-capitalization indexes, in which a stock is  replaced when its capitalization falls to low. 

The {\em numeraire portfolio}, or {\em log-optimal portfolio}, is that portfolio with the highest expected logarithmic growth rate at all times.  A {\em numeraire market} is a stock market for which the numeraire portfolio is the market portfolio. It was shown in \citeN{KK:2018} that a  closed market that is also a numeraire market will be asymptotically unstable, with all the capital concentrating into a single stock. Accordingly, we shall concentrate here on open markets. We first need to review some basic definitions and elementary results regarding stock-market models. For clarity of exposition, we shall first consider closed markets and postpone consideration of open markets until a later section.

Suppose that a market defined on the time interval $[0,T]$ is composed of stocks $X_1,\ldots,X_n$, where $n>1$ and the $X_i$ are represented by positive \ito\ processes that satisfy
\begin{equation}\label{1}
d\log X_i(t) = \g_i(t)\,dt + \sum_{\ell=1}^d\xi_{i\ell}(t)\,dW_\ell(t),
\end{equation}
for $t\in[0,T]$, where $(W_1,\ldots,W_d)$ is an $d$-dimensional Brownian motion for $d\ge n$, and $\g_i$ and $\xi_{_i\ell}$ are measurable functions adapted to the Brownian filtration. We shall assume throughout that the stochastic differential equations we encounter have at least weak solutions. The process $\g_i$ is called the $i$th {\em growth rate process}. We define the {\em rank process}  to be the random permutation $r_t\in\Sigma_n$ such that $r_t(i)<r_t(j)$ if $X_i(t)>X_j(t)$ or if $X_i(t)=X_j(t)$ and $i<j$. We also define the {\em ranked stock processes} $X_{(1)},\ldots,X_{(n)}$ such that $X_{(r_t(i))}(t)=X_i(t)$, so  $X_{(1)}(t)\ge\cdots\ge X_{(n)}(t)$. We can also define the inverse permutation $p_t\eqdef r_t^{-1}\in\Sigma_n$, so we have $X_{p_t(k)}(t)=X_{(k)}(t)$.

From \eqref{1} we can derive the {\em variance rate processes} $\s^2_i$ for $i=1,\ldots,n$ and {\em covariance rate processes} $\s_{ij}$ for $i,j=1,\ldots,n$,
with 
\[
\s^2_i(t)\eqdef \frac{d}{dt}\brac{\log X_i}_t=\sum_{\ell=1}^d \xi^2_{i\ell}(t)\quad\text{ and }\quad\s_{ij}(t)\eqdef\frac{d}{dt}\brac{\log X_i,\log X_j}_t= \sum_{\ell=1}^d \xi_{i\ell}(t)\xi_{j\ell}(t),
\]
for $t\in[0,T]$. 
We shall also have an additional process $X_0$, called {\em cash,} which has no stochastic term, so
\[
d\log X_0(t) = \g_0(t)\,dt,
\]
for $t\in[0,T]$. For cash the quadratic variation processes  $\brac{X_0}_t\equiv 0$ and $\brac{X_0,X_i}_t\equiv 0$, for $i=1,\dots,n$. Since by subtracting the growth rate $\g_0$  from all the growth rates $\g_i$ we can change the frame of reference to give cash a zero growth rate, there is no loss of generality in assuming that $X_0\equiv1$ and $\g_0\equiv 0$, and we shall do so. 

For each stock $X_i$ we define the {\em rate of return process} $\a_i$  by
\[
\a_i(t)\eqdef \g_i(t)+\frac{\s^2_i(t)}{2},
\]
and for cash, $\a_0\equiv 0$. In this case, \ito's rule implies that
\[
\frac{dX_i(t)}{X_i(t)} = \a_i(t)\,dt +  \sum_{\ell=1}^d\xi_{i\ell}(t)\,dW_\ell(t),\as
\]
For background on this material in this section, we refer the reader to \citeN{F:2002} and  \citeN{FK:2009}.

A (generalized) {\em portfolio} $\p$ is a family of bounded measurable processes $\p_1,\ldots,\p_n$ that are adapted to the Brownian filtration. For each portfolio $\p$ there is an implicit {\em cash weight} $\p_0$, with
\[
\p_0(t)\eqdef 1 -\p_1(t)-\cdots-\p_n(t),
\]
so that
\[
\sum_{i=0}^n \p_i(t)=1,
\]
for all $t\in[0,T]$. The value process $Z_\p$ of the portfolio $\p$  satisfies
\begin{align*}
\frac{d Z_\p(t)}{Z_\p(t)}&=\sumi \p_i(t) \frac{dX_i(t)}{X_i(t)}\\
&=\sumi \p_i(t) \a_i(t)\,dt + \bigg(\sumij \p_i(t) \p_j(t)\s_{ij}(t)\bigg)^{1/2}dW(t),\as,
\end{align*}
for some Brownian motion $W$. (Cash can be ignored since it contributes nothing to the return of the portfolio.) From this we can define the {\em portfolio rate of return process} $\a_\p$ by
\[
\a_\p(t)\eqdef \sumi \p_i(t) \a_i(t),
\]
and the {\em portfolio variance rate process} $\s^2_\p$ by
\[
\s^2_\p(t)\eqdef\sumij \p_i(t) \p_j(t)\s_{ij}(t),
\]
in which case
\begin{equation}\label{3}
\frac{dZ_\p(t)}{Z_\p(t)} = \a_\p(t)\,dt +  \s_\p(t)\,dW(t),\as
\end{equation}
By an application of \ito's rule we can derive the logarithmic representation for $\p$, with 
\begin{equation}\label{4}
d\log Z_\p(t) = \g_\p(t)\,dt +  \s_\p(t)\,dW(t),\as,
\end{equation}
where the portfolio growth rate satisfies
\[
\g_\p(t)=\a_\p(t)-\frac{\s^2_\p(t)}{2},\as
\]

The {\em market portfolio process} $X$ is defined by
\[
X(t)\eqdef X_1(t)+\cdots+X_n(t),
\]
with no cash in the market portfolio. The corresponding {\em market weight processes} $\m_i$ are defined by
\[
\m_i(t)\eqdef\frac{X_i(t)}{X(t)},
\]
for $i=1.\ldots,n$, so $\m_1(t)+\cdots+\m_n(t)=1$. It can be shown that the portfolio value process $Z_\m$ corresponding to the market wights satisfies
\begin{equation}\label{4.1}
Z_\m(t)=X(t),\as,
\end{equation}
for $t\in[0,T]$, if $Z_\m(0)=X(0)$.

\section{ Numeraire markets}%%%%%%%%%%%%%%%%%%%%%%%%%

The {\em numeraire portfolio} $\nu$ is the portfolio with the highest portfolio growth rate, so for any portfolio $\p$, $\g_\nu(t)\ge\g_\p(t)$ for $t\in[0,T]$.
Since
\begin{align}
\g_\p(t)&=\a_\p(t)-\frac{\s^2_\p(t)}{2}\notag\\
&=\ppi^T(t)\aalpha(t)-\half\ppi^T(t)\SSigma(t)\ppi(t),\as,\label{5}
\end{align}
where $\aalpha(t)=(\a_1(t),\ldots,\a_n(t))^T$, $\ppi(t)=(\p_1(t),\ldots,\p_n(t))^T$, and $\SSigma(t)$ is the covariance matrix with entries $\s_{ij}(t)$. To maximize the growth rate at time $t$, we find the unconstrained maximum for  $\g_\p(t)$, which occurs  when
\begin{align*}
\nabla\g_\p(t)&= \aalpha(t) - \SSigma(t) \ppi(t)\\
&=0,\as,
\end{align*}
where $\nabla=(D_1,\ldots,D_n)^T$ is the gradient with respect to $\ppi(t)$. Hence, for the numeraire portfolio $\nu$ we have 
\[
\aalpha(t) = \SSigma(t) \nnu(t),\as,
\]
where $\nnu(t)=(\nu_1(t),\ldots,\nu_n(t))^T$. This is equivalent to
\begin{align}
\a_i(t)&=\sumj \nu_j(t)\s_{ij}(t)\notag\\
&=\s_{i\nu}(t),\as,\label{6}
\end{align}
for $i=1,\ldots,n$. The process $\s_{i\nu}$ is called the {\em portfolio covariance process} for $Z_\nu$ and $X_i$.

For a numeraire market, the numeraire portfolio $\nu$ is the same as the market portfolio $\m$, so \eqref{6} becomes
\[
\a_i(t)=\s_{i\m}(t),\as,
\]
which, in terms of growth rates, is
\begin{equation}\label{7}
\g_i(t)=\s_{i\m}(t)-\frac{\s^2_i(t)}{2},\as
\end{equation}

\section{Open markets}%%%%%%%%%%%%%%%%%%%%%%%%%

Consider an equity universe $X$ of the form \eqref{1} with stock price processes $X_1,\ldots,X_N$, with $N>2$, and {\em universal weight processes} $\z_1,\ldots,\z_N$ defined by
\[
\z_i(t)\eqdef\frac{X_i(t)}{X_1(t)+\cdots+X_N(t)}.
\]
  The {\em open market} $X^*_n\subset X$ of size $n<N$ is the subset of stocks that occupy the top $n$ ranks at any given time. The {\em market capitalization process} $X_{[n]}$ for the market $X^*_n$  is defined by
\begin{equation}\label{100}
X_{[n]}(t)\eqdef X_{(1)}(t)+\cdots+X_{(n)}(t),
\end{equation}
and the market portfolio $\m$ for $X^*_n$  is the portfolio in $X$ defined by the weights $\m_1,\ldots,\m_N$, where
\begin{equation*}
\m_{p_t(k)}(t)=\begin{cases}\displaystyle \frac{X_{(k)}(t)}{X_{[n]}(t)},\quad &k=1,\ldots,n\\
0,  &k=n+1,\ldots,N.
\end{cases}
\end{equation*}
The ranked market weight processes $\m_{(k)}$ satisfy
\[
\m_{(k)}(t)=\frac{X_{(k)}(t)}{X_{[n]}(t)},\as,
\]
for $k=1,\ldots,n$.

Let  $\lt{k,k+1}$ be the local time for $\log (X_{(k)}/X_{(k+1)})$ at the origin, for $k=1,\ldots,N-1$.  The local time $\lt{k,k+1}$ is equal to the local time at the origin  for $\log (\z_{(k)}/\z_{(k+1)})$ and for $\log (\m_{(k)}/\m_{(k+1)})$. From Example~4.3.2 of \citeN{F:2002} or Lemma~2.1 of \citeN{FF:2017} we see that the dynamics of the market portfolio value process $Z_\m$ will follow
\begin{equation}\label{8}
d \log Z_\m(t) = d\log X_{[n]}(t) - \half \m_{(n)}(t)d\lt{n,n+1}(t),\as
\end{equation}
The appearance of local time in \eqref{8} is similar to the leakage from functionally generated portfolios in Example~4.3.5 of \citeN{F:2002}, and this shows that the market capitalization process grows at a faster rate than the market portfolio.
 
\defn{  (\citeN{F:2002}) A system of positive continuous semimartingales $X_1, \ldots, X_N$ is {\em asymptotically stable} if
\begin{enumerate}
\item $\displaystyle
       \lim_{t\to\infty}t^{-1}\log (X_i(t)/X_j(t))=0,\as$, for $i,j=1,\ldots,N$ ({\em coherence});
\item $\displaystyle
      \lim_{t\to\infty}t^{-1}\lt{k,k+1}(t)\eqdef\bla_{k,k+1}>0,\as$, for $k=1,\ldots,N -1$;
\item $\displaystyle
        \lim_{t\to\infty}t^{-1}\brac{\log (X_{(k)}/X_{(k+1)})}_t \eqdef\bsi^2_{k,k+1}>0,\as$, for $k=1,\ldots,N -1$.
\end{enumerate}
}

\rem{ {\em CAPM-consistent markets.} In the classical portfolio theory of \citeANP{Markowitz:1952} \citeyear{Markowitz:1952,Markowitz:1959}, a  portfolio $\p$ is defined to be {\em efficient} if it has  the minimum portfolio variance $\s^2_\p(t)$ for its expected return $\a_\p(t)$. The celebrated capital asset pricing model (CAPM) of \citeN{Sharpe:1964} concluded that if all investors hold efficient portfolios, then the market portfolio $\m$ will also be efficient, and all efficient portfolios will be a combination of the market portfolio and cash. The numeraire portfolio is an efficient portfolio, since if another portfolio had the same expected return but lower variance, that portfolio would have a higher growth rate. Hence, under CAPM, the numeraire portfolio will be a combination of cash and the market portfolio $\nu$. In this case, \eqref{7} becomes
\begin{equation}\label{7.1}
\g_i(t)=\rho_\nu(t)\s_{i\m}(t)-\frac{\s^2_i(t)}{2},\as,
\end{equation}
where $\rho_\nu(t)$ is the proportion of market portfolio $\m$ held in the numeraire portfolio $\nu$ at time $t$, and $(1-\rho_\nu(t))$ is the proportion of cash. A market in which \eqref{7.1} is satisfied is called a {\em CAPM-consistent market.} Since condition \eqref{7} is stronger than condition \eqref{7.1}, a numeraire market is also a CAPM-consistent market. However, closed numeraire markets are asymptotically unstable (see \citeN{KK:2018}), while examples of closed, asymptotically stable,  CAPM-consistent markets can be constructed in a manner similar to that used below for the construction of numeraire portfolios in open markets. Nevertheless, examples of closed, CAPM-consistent markets seem to bear little resemblance to real stock markets.} 

\section{An example of a numeraire market}%%%%%%%%%%%%%%%%%%%%%%%%%

We wish to construct a model of a numeraire market that is asymptotically stable, and since we know that first-order models are asymptotically stable, we shall start with an equity universe that has a  structure similar to that of a first-order model (see, e.g., \citeN{F:2002} of \shortciteN{BFK:2005}).

 Let us consider the equity universe $X$ defined by
\begin{equation}\label{9}
d\log X_i(t) = \big(g_{r_t(i)}(t)+G\big)dt+ s_{r_t(i)}dW_i(t)+S\,dW(t),\as,
\end{equation}
for $i=1,\ldots,N$, with $N\ge 2$,  a constant $G$, a constant $S^2\ge0$,  positive constants $s^2_1,\ldots,s^2_N$, an $N+1$-dimensional  Brownian motion $(W_1,\ldots,W_N,W)$, and processes $g_1,\ldots,g_N$ defined by 
\[
g_k(t)=\ph(k,\z_{(1)}(t),\ldots,\z_{(N)}(t))>0,
\]
where $\ph$ is a  bounded  continuous function such that for some $\eps>0$,
\begin{equation}\label{10}
g_1(t)+\cdots+g_N(t)=0 \quad\text{ and }\quad g_1(t)+\cdots+g_k(t)<-\eps,\quad\text{ for }\quad 1\le k<N.
\end{equation}

The model \eqref{9} is quite similar to the first-order models  in \shortciteN{BFK:2005} and and hybrid Atlas models in \shortciteN{IPBKF:2011}, and results for these models should generally remain valid in our current setting. We see that \eqref{9} is Markovian with a.s.\ continuous sample paths, and the eigenvalues of the covariance matrix are bounded away from zero.  Hence, we can follow the reasoning of \shortciteN{IPBKF:2011}, Theorem~1, to establish the existence of a stable distribution for the gap processes $\log(X_{(k)}/X_{k+1)})$ (see also Theorems~4.1 and 5.1 of \citeN{Khasminskii:1980}). Then the proof of Lemma~1 of \shortciteN{IPBKF:2011} can be used to show that the system will spend zero local time at triple points. Since the $g_k$ are all defined by the same function $\ph$, the processes $X_i$ will be exchangeable and hence they will asymptotically spend equal time at each rank, as is the case for first-order models in \shortciteN{BFK:2005}. Therefore, all the $X_i$ will asymptotically share the average growth rate $G$, and this will be the growth rate of the universe, so the system will be coherent.  The analysis of \shortciteN{IPBKF:2011} for hybrid Atlas models remains valid here, so the system \eqref{9} will be  asymptotically stable. 

Now, for some $n<N$ let us consider the open submarket $X^*_n\subset X$ with market portfolio process $\m$.  Then 
\begin{equation}\label{11.1}
\m_i(t)=\1_{\{r_t(i)\le n\}}\frac{X_i(t)}{X_{[n]}(t)},\quad i=1,\ldots,N,\quad\text{ and }\quad\m_{(k)}(t)=\1_{\{k\le n\}}\frac{X_{(k)}(t)}{X_{[n]}(t)},\quad k=1,\ldots,N.
\end{equation}
It is convenient to extend the definition of these weights to
\[
\mtil_i(t)\eqdef\frac{X_i(t)}{X_{[n]}(t)},\quad i=1,\ldots,N,
\]
so we have $\m_i(t)=\mtil_i(t)$ for $r_t(i)\le n$ and $\m_{(k)}(t)=\mtil_{(k)}(t)$ for $k\le n$. With this notation, the submarket $X^*_n$ defined by \eqref{9} will satisfy  \eqref{7} if
\begin{equation}\label{11.2}
g_{r_t(i)}(t)+G=\m_i(t)s^2_{r_t(i)}+S^2-\half\big(s^2_{r_t(i)}+S^2\big),\as,
\end{equation}
for $i=1,\ldots,n$. Hence,  if the processes $g_k$ defined by
\begin{equation}\label{11}
g_k(t)=\Big(\mtil_{(k)}(t)-\half\Big)s^2_k-G +\half S^2,
\end{equation}
are a.s.\ negative for $k=1,\ldots,N-1$, and
\begin{equation}\label{12}
g_{N}(t)=-\big(g_1(t)+\ldots+g_{N-1}(t)\big),
\end{equation}
then \eqref{7} as well as \eqref{10} will be satisfied and $X^*_n$ will be a numeraire market.
Accordingly, our strategy to construct a numeraire market will be to find positive constants $s^2_1,\ldots,s^2_N$ and constants $G$ and $S^2\ge0$ such that the $g_k(t)$ defined by \eqref{11} satisfy condition \eqref{10}. 

\section{ A simulated numeraire market}%%%%%%%%%%%%%%%%%%%%%%%%%

Here we shall construct a numeraire market of the form \eqref{9} following the strategy we just outlined. Let 
\begin{equation}\label{13}\begin{split}
n&=500,\\
N&=550,\\
s^2_k&=0.06, \quad k=1,\ldots,N,\\
S^2&=0.04,\\
G&=0.051,
\end{split}\end{equation}
with
\begin{equation}\label{14}\begin{split}
g_k(t)&=0.06\Big(\mtil_{(k)}(t)-\half\Big)-0.031<-0.001,\quad k=1,\ldots,N-1,\\
g_{N}(t)&=-(g_1(t)+\cdots+g_{N-1}(t)),
\end{split}\end{equation}
which corresponds to  \eqref{11}.  We see from \eqref{14} that $g_k(t)<0$ for $k=1,\ldots,N-1$, so condition \eqref{10} is satisfied.

The results of a simulation of this market are presented below in Figures~\ref{f1} and \ref{f2}. The simulation was run over 2~million iterations, with the first million used for convergence and the second million recorded in the charts. Figure~\ref{f1} shows the average distribution of the simulated rank-based market weights, represented by the black curve, along with the distribution curve for the first-order approximation represented by the red curve. The first-order approximation, defined in Section~5.5 of \citeN{F:2002}, uses the estimate
\[
\limT{1}\intT\log\big(X_{(k)}(t)/X_{(k+1)}(t)\big)dt=\frac{\bsi^2_{k,k+1}}{2\bla_{k,k+1}},\as,
\]
which holds if stable distribution for the gap processes $\log(X_{(k)}/X_{(k+1)})$ is exponential (see \shortciteN{BFK:2005}). The closeness of  the two curves suggests  that the stable distribution for the gap processes is close to exponential.   The $\bla_{k,k+1}$ were estimated from the simulation using \eqref{8}, and the $\bsi^2_{k,k+1}=s^2_k+s^2_{k+1}$ came directly from the parameter definitions.

Figure~2 shows the cumulative logarithmic growth of the market portfolio and the market capitalization for the simulated numeraire market. We see that  the market capitalization $X_{[n]}$ grows at a faster rate than the value process $Z_\m$ of the market portfolio, as we expected from \eqref{8}.
 
The R code used for this simulation is included below.

\rem{ The model with parameters \eqref{13} is somewhat simpler than \eqref{9}. Indeed, we can write
\begin{equation}\label{15}
d\log X_i(t) = \big(g_i(t) +G-\1_{\{r_t(i)=N\}}g(t)\big)dt+ \s\,dW_i(t)+S\,dW(t),\as,
\end{equation}
for $i=1,\ldots,N$, where the Brownian motions $W_i$ and $W$ are as in \eqref{9}, the parameters are
\begin{equation*}\begin{split}
n&=500,\\
N&=550,\\
\s^2&=0.06,\\
S^2&=0.04,\\
G&=0.051,
\end{split}\end{equation*}
and where
\[
g_i(t)=\mtil_i(t)\s^2-G +\half \big(S^2-\s^2\big)<-0.001,
\]
with 
\[
g(t)=g_1(t)+\cdots+g_N(t).
\]
Since the processes
\[
dY_i(t)=d\log X_i(t)-S\,dW(t),
\]
 for $i=1,\ldots,N$, are independent and all have the same variance rate $\s^2$,  Girsanov's theorem (see \citeN{Karatzas/Shreve:1991}) implies that $Y_1,\ldots,Y_N$ will be equivalent to  $N$-dimensional Brownian motion. It follows that the processes $Y_i$ will have no triple points, so the same will hold for the processes $X_i$. Let us note also that the parameter processes $\g_i$ and $\s^2_i$ for the stocks $X_i$ in the numeraire market $X^*_n$ do not depend on rank. }

\section{Conclusion}%%%%%%%%%%%%%%%%%%%%%%%%%

We have  constructed an example of an asymptotically stable open numeraire market and considered some of its basic characteristics. We have seen that the asymptotic behavior of open numeraire markets is similar to that of markets comprising the top ranks of first-order models or hybrid Atlas models. For an asymptotically stable open numeraire market, over the long term no portfolio strategy will dominate the market portfolio, and the market capitalization process will dominate all portfolio strategies.

%\pagebreak
\bibliographystyle{chicago}
\bibliography{math}

\pagebreak
\vspace*{-85pt}
\begin{figure}[H]
\begin{center}
\hspace*{-20pt}
\scalebox{.80}{ \rotatebox{0}{
\includegraphics{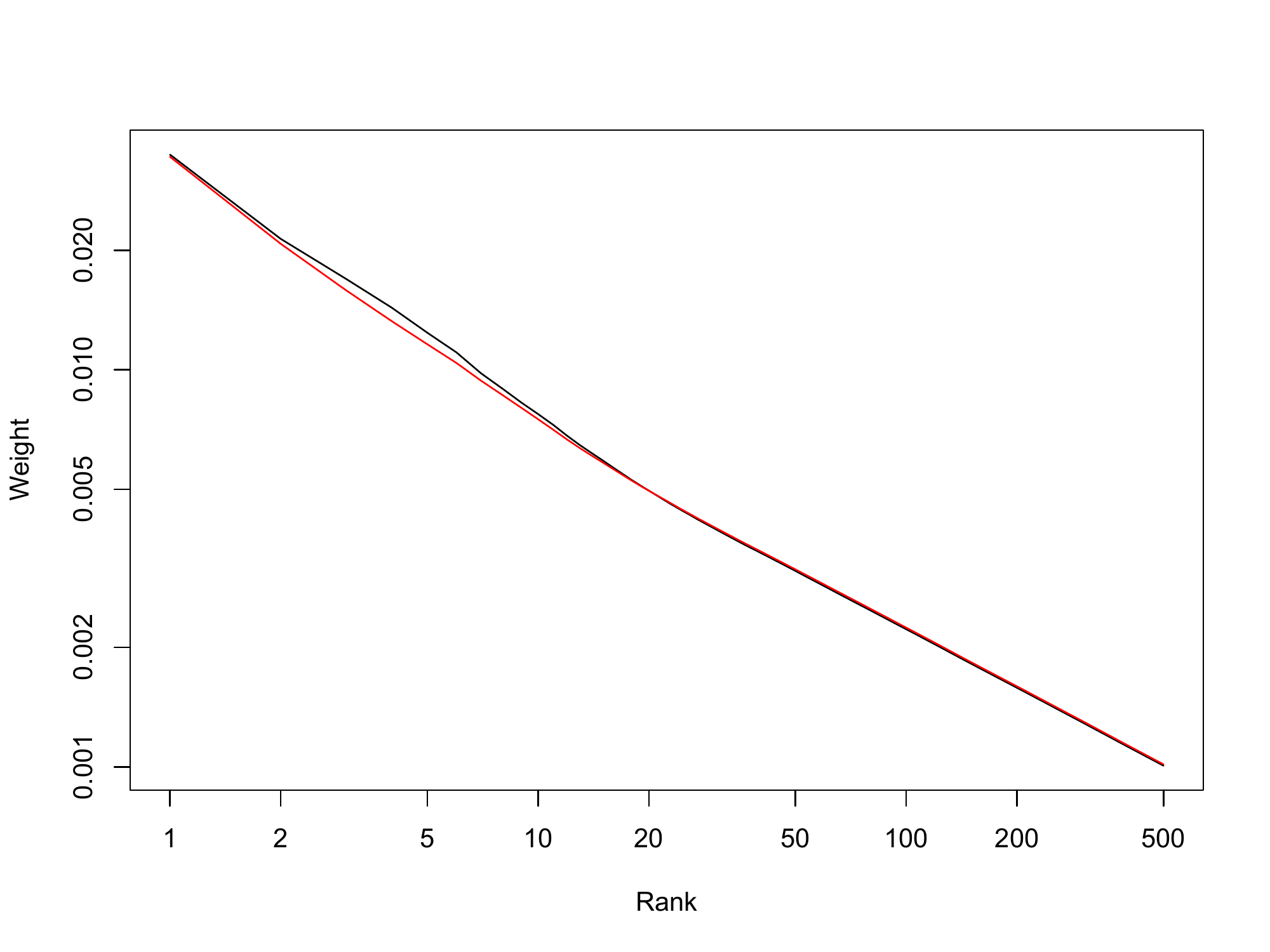}}}
\vspace{-20pt}
\caption{Weight distributions related to the open market \eqref{9} with parameters \eqref{13} and \eqref{14}.}\label{f1}
Simulated; black. First-order; red.
\end{center}
\end{figure}

\vspace*{-60pt}
\begin{figure}[H]
\begin{center}
\hspace*{-20pt}
\scalebox{.80}{ \rotatebox{0}{
\includegraphics{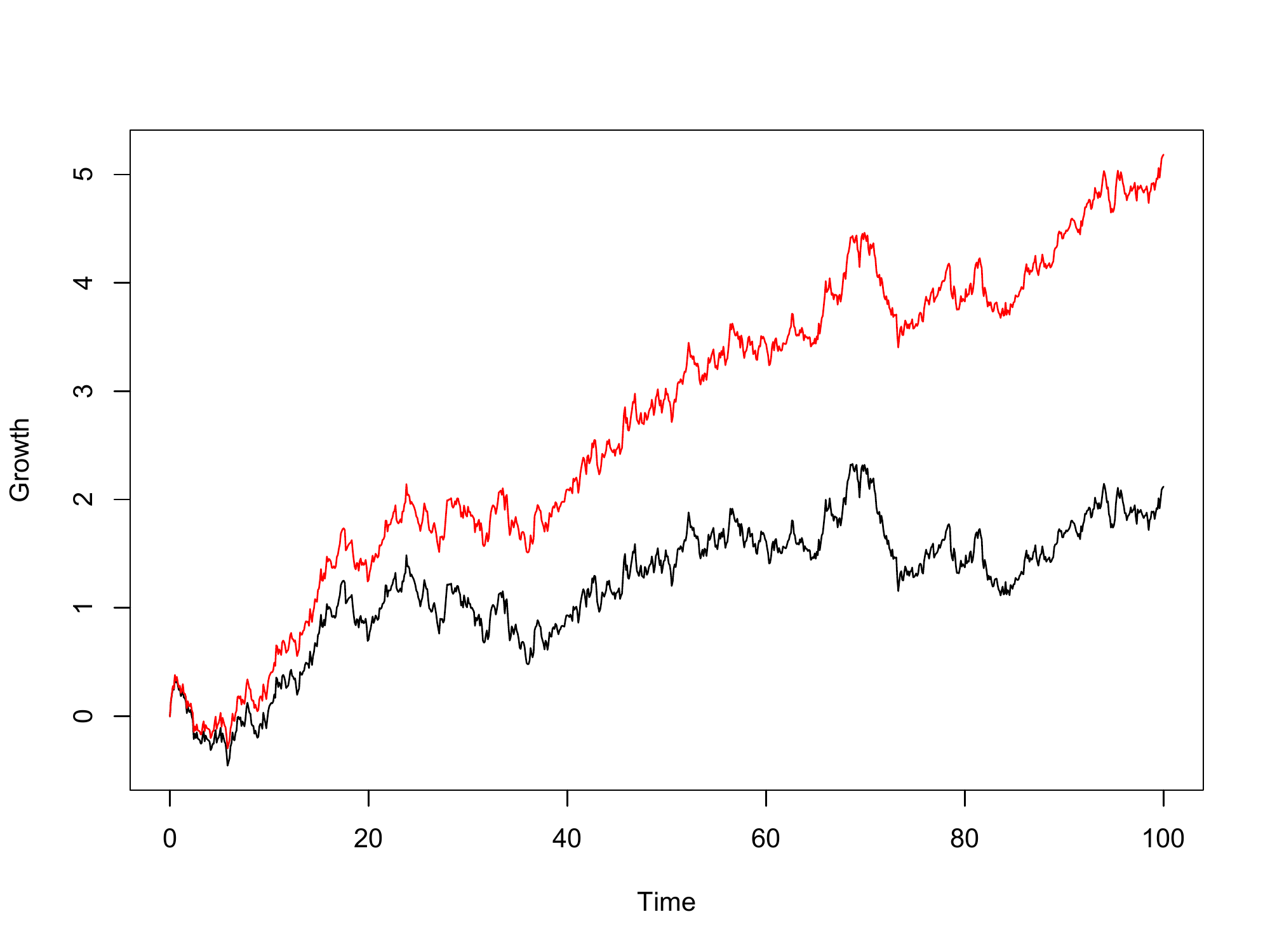}}}
\vspace{-20pt}
\caption{Cumulative log-growth of the market portfolio value process $Z_\m$, from \eqref{8}; black.}\label{f2}
Cumulative log-growth of the market capitalization process $X_{[n]}$, from \eqref{100}; red.
\end{center}
\end{figure}

\pagebreak
\vspace{10pt}
\centerline{\Large\bf R code for the simulation}   %  numark5.txt
\vspace{10pt}
\begin{Verbatim} 
n<-500
N<-550
T<-1000000
G0<-  .051
SS0<- .04
ss1<- .06
ss2<-.06
Tfac=100/T
G<-Tfac*G0
if (ss1 == ss2)ss<-rep(ss1,N) else ss<-seq(ss1,ss2,(ss2-ss1)/(N-1))
ss<-Tfac*ss
SS<-Tfac*SS0
S<-sqrt(SS)
s<-sqrt(ss)
x<- -log(1:N)
x<-x-log(sum(exp(x)))
xx<-exp(x)
z<-rep(1,N)
g<-rep(0,N)
gg<-rep(0,N)
LT<-rep(0,N)
GR<-NULL
GG<-NULL
T1<-2*T
tt<-0
for(t in 1:T1){
	mu<-xx/sum(xx[1:n])
	g<-(.5-mu)*ss+G-.5*SS
	XX<-xx
	ng2<-sum(g)/2
	x<-x-g+s*rnorm(N)
	X1<-x
	GS<-G+S*rnorm(1)
	xx0<-log(sum(exp(x+GS)))
	ii<-order(x)
	x0<-0
	for(k in 1:(N-1)){
		x0<-x0+x[ii[k]]
		if(k*x[ii[k+1]]-x0 > ng2)break
	}
	xpiv<-2*(x0+ng2)/k
	x[ii[k:1]]<-xpiv-x[ii[1:k]]
	x<- -sort(-x)
	xx<-exp(x+GS)
	yy<-sum(xx)
	xx<-xx/yy
	x<-log(xx)
	if(t%%(T%/%10) == 0)print(c(t,k,xx[1]))
	if(t > T1-T){
		tt<-tt+1
		gg<-gg+g
		GG[tt]<-log(yy)
		GR[tt]<-xx0
		z<-z+xx 
		X2<-cumsum(exp(-sort(-X1)))
		X1<-cumsum(exp(X1))
		LT<-LT+2*(X2-X1)/XX
	}
}
z<-z[1:n]/sum(z[1:n])
gg<-gg/tt
lt<-2*cumsum(gg)
LT<-LT/T
for(i in 1:n)if(LT[i] <= 0){
	print(c(i,LT[i],lt[i]))
	LT[i]<-lt[i]
}
Time=seq(0,tt,tt/1000)
Time[1]<-1
GR<-cumsum(GR)[Time]
GG<-cumsum(GG)[Time]
ZZ<-ss/LT
ZZ<-exp(cumsum(ZZ[n:1]))[n:1]
ZZ<-ZZ/sum(ZZ)
graphics.off()
quartz('1',width=8,height=6)
plot(Time*Tfac,GR,ylim=range(GR,GG),type='l',xlab='Time',ylab='Growth')
lines(Time*Tfac,GG,col=2)
quartz('2',width=8,height=6)
plot(-gg/Tfac,type='l',xlab='Rank',ylab='g_k')
quartz('3',width=8,height=6)
plot(z,type='l',ylim=range(z,ZZ),log='xy',xlab='Rank',ylab='Weight')
lines(ZZ,col=2)
\end{Verbatim}

\end{document}